\documentclass[11pt,a4paper]{article}
\usepackage{amsmath}
\usepackage[mathcal]{eucal}
\usepackage{latexsym}
\usepackage{amssymb}

\begin{titlepage} 
\title{Covariance of the algebra of constraints of 4-D general relativity}
\author{Eyo Eyo Ita III}
\medskip
\input amssym.def
\input amssym.tex

\def \in{\indent}

\begin{document}
\maketitle
\bigskip
\centerline{Department of Applied Mathematics and Theoretical Physics} 
\smallskip
\centerline{Centre for Mathematical Sciences, University of Cambridge, Wilberforce Road}
\smallskip
\centerline{Cambridge CB3 0WA, United Kingdom}
\smallskip
\centerline{eei20@cam.ac.uk} 
    
\bigskip
  
\begin{abstract}
We argue that the standard canonical treatment of GR breaks manifest spacetime covariance.  We present new variables which carry a reducible representation of gauge transformations and spacetime diffeomorphisms.  A proposal is presented for an action designed to realize these symmetries at the canonical level.
\end{abstract}
\end{titlepage}

\section{Introduction}

In the canonical formulation of general relativity (GR) one performs a 3+1 splitting into spatial variables which evolve in time.  This seems to go directly against the principle of general covariance, where space and time manifestly appear on an equal footing.  In this paper we will show for certain fomulations of general relativity that this covariance becomes manifestly broken at the canonical level.  Our aim in this paper will be to establish that a certain field provides an off-shell realization of the algebra of spacetime diffeomorphisms, and we will propose an action for gravity based on this field, with a view toward restoration of the original covariance of the theory.  The full test of such a restoration at the canonical level will be reserved for future research.\par
\indent  
Consider a general transformation of coordinates $x\in{M}$, where $M$ is a 4-dimensional spacetime manifold
\begin{eqnarray}
\label{GENCOORD}
x^{\mu}\rightarrow{x^{\prime}}^{\mu}={x}^{\mu}+\xi^{\mu}(x).
\end{eqnarray}
\noindent
The transformations (\ref{GENCOORD}) induce the following Lie bracket between vector fields $\xi,\zeta\in{M}$
\begin{eqnarray}
\label{LIE}
\bigl[\xi^{\mu}\partial_{\mu},\zeta^{\nu}\partial_{\nu}\bigr]=\bigl(\xi^{\mu}\partial_{\mu}\zeta^{\nu}-\zeta^{\mu}\partial_{\mu}\xi^{\nu}\bigr)\partial_{\nu},
\end{eqnarray}
\noindent
which we will for the purposes of this paper regard as a Lie algebra of general coordinate transformations.\par
\indent
We would like find a description of GR which realizes this algebra at the canonical level.  For later comparison let us perform a decomposition of (\ref{LIE}) into spatial and temporal parts.  For the commutator between two spatial diffeomorphisms we should choose $\xi^{\mu}=\delta^{\mu}_iM^i$ and $\zeta^{\mu}=\delta^{\mu}_jM^j$.  The corresponding algebra is given by
\begin{eqnarray}
\label{LIE3}
\bigl[N^i\partial_i,N^j\partial_j\bigr]=\bigl(M^i\partial_iN^j-N^i\partial_iM^j\bigr)\partial_j.
\end{eqnarray}
\noindent
The commutator of two spatial diffeomorphisms is a spatial diffeomorphism, which shows that at the level of the vector fields, spatial diffeomorphisms should form a subalgebra of (\ref{LIE}).  Moving on to the commutator of a spatial with a temporal diffeomorphism, we make the identifications $\xi^{\mu}=\delta^{\mu}_iN^i$ and $\xi^{\mu}=\delta^{\mu}_0N$, yielding
\begin{eqnarray}
\label{LIE4}
\bigl[N^i\partial_i,N\partial_0\bigr]=(N^i\partial_iN)\partial_0-(N\dot{N}^i)\partial_i.
\end{eqnarray}
\noindent
The commutator of a spatial with a temporal diffeomorphism is a linear combination of the two transformations.\par
\indent  
Moving on to the commutator of two temporal diffeomorphisms, we choose $\xi^{\mu}=\delta^{\mu}_0M$ 
and $\zeta^{\mu}=\delta^{\mu}_0N$, which yields
\begin{eqnarray}
\label{LIE5}
\bigl[M\partial_0,N\partial_0\bigr]=(M\dot{N}-N\dot{M})\partial_0.
\end{eqnarray}
\noindent
The commutator of two temporal diffeomorphisms is another temporal diffeomorphism, constituting a subalgebra of (\ref{LIE}).  This suggests for a theory invariant under general coordinate transformations, that the fully covariant algebra of spacetime diffeomorphisms should decompose into two subalgebras, a spatial one and a temporal one.\par
\indent  
The Poisson algebra of hypersurface deformations has been computed by Teitelboim \cite{TEITEL}
\begin{eqnarray}
\label{ALGEBRA121}
\{\vec{H}[\vec{N}],\vec{H}[\vec{M}]\}=H_k\bigl[N^{i}\partial^{k}M_i-M^{i}\partial^{k}N_i\bigr];\nonumber\\
\{H[{N}],\vec{H}[\vec{N}]\}=H[N^{i}\partial_{i}{N}\bigr]\nonumber\\
\bigl[H[{N}],H[{M}]\bigr]
=H_{i}[\bigl({N}\partial_{j}{M}-{M}\partial_{j}{N}\bigr)q^{ij}],
\end{eqnarray}
\noindent
where $H^{\mu}=(H,H_i)$ refer to the Hamiltonian and diffeomorphism constraints, smeared by a lapse function and shift vector $N^{\mu}=(N,N^i)$.  If $H$ and $H_i$ generate spatial and temporal diffeomorphisms, then one would like to be able to make the identifications
\begin{eqnarray}
\label{WEWOULDLIKE}
H\longrightarrow\partial_0;~~H_i\longrightarrow\partial_i.
\end{eqnarray}
\noindent
Comparison of (\ref{ALGEBRA121}) with (\ref{LIE3}), (\ref{LIE4}) and (\ref{LIE5}) shows some similarities and some differences.  The part of (\ref{ALGEBRA121}) involving purely spatial diffeomorphisms is isomorphic 
with (\ref{LIE3}).  However, the second equation of (\ref{ALGEBRA121}) is isomorphic with (\ref{LIE4}) only when $\dot{N}^i=0$, which means that the shift vector $N^i$ must be independent of time.  Additionally, aside from the issue of phase space structure functions $q^{ij}$, the third line of (\ref{ALGEBRA121}) is not isomorphic with (\ref{LIE5}).  Since two Hamiltonian constraints Poisson-commute into a diffeomorphism (and not a Hamiltonian) constraint, then the interpretation is that the algebra (\ref{ALGEBRA121}) breaks manifest 4-dimensional diffeomorphism covariance.\par
\indent  
There are various nonmetric formulations of GR where the basic variables include a gauge connection $A^a_{\mu}$ (See e.g. \cite{SPINCON}, \cite{ASH1}).  In this paper we will focus on the Ashtekar formulation \cite{ASH1}, demonstrating that even in this case there are similar discrepancies with (\ref{LIE}) regarding the temporal parts of the Poisson algebra of constraints.  With a view toward restoring this algebra to be more in line with (\ref{LIE}), we will propose a new action for GR, called the instanton representation $I_{Inst}$.  This paper will show that a certain subset of the variables of $I_{Inst}$ forms an off-shell representation of the algebra (\ref{LIE}), reserving the check for its full canonical realization for a separate paper.\par
\indent
The organization of this paper is as follows.  Section 2 will establish the symmetry group of a 4-D gauge connection $A^a_{\mu}$ as $SO(3,C)*Diff$.  Our notation signifies that this symmetry is an off-shell symmetry, derived independently of any equations of motion, Poisson brackets or canonical structure.  Section 3 brings in the Ashtekar formulation which uses $A^a_{\mu}$ as one of the dynamical variables.  We show that the original $SO(3,C)*Diff$ symmetry of $A^a_{\mu}$ is not realized at the canonical level.  Section 4 proposes a new variable $\Psi_{ae}$, which in conjunction with $A^a_{\mu}$ forms the basis for our proposition.  This proposition is provided in the discussion section in the form of an action $I_{Inst}[A,\Psi]$.  We demonstrate 
that $\Psi_{ae}$ preserves at least the spatial part of the $SO(3,C)*Diff$ algebra, relegating the temporal part for future research.  The temporal part becomes of interest when one wishes to obtain a reduced phase space 
under $SO(3,C)*diff$, the spatial part of the fully covariant algebra.\par
\indent
On a final note regarding index conventions, symbols from the beginning of the Latin alphabet $a,b,c,\dots$ will denote internal $SO(3,C)$ indices, while from the middle $i,j,k,\dots$ will denote spatial indices corresponding to a 3-D spatial manifold $\Sigma$.  Spacetime indices will be denoted by Greek symbols $\mu,\nu\dots$, and in the canonical decomposition we will assume a spacetime of topology $M=\Sigma\times{R}$, where $M$ is a globally hyperbolic manifold of 4-D spacetime.

\section{Symmetry group of gauge connections}

The form variation of a 4-dimensional $SO(3,C)$ gauge connection $A^a_{\mu}=(A^a_0,A^a_i)$ under gauge transformations and spacetime diffeomorphisms will be established in this section.  We will show that $A^a_{\mu}$ forms an off-shell representation of the $SO(3,C)*Diff$ group.  By this we mean that the algebra for this group closes at the covariant level independently of any equations of motion or symplectic structure.  For notational purposes for this paper, we will distinguish the algebra $SO(3,C)*diff$ from $SO(3,C)*Diff$.  In this notation the latter refers to full spacetime diffeomorphisms while the former refers to only the spatial parts of these diffeomorphisms, which excludes the temporal diffeomorphisms.\par
\indent  
Under an infinitesimal $SO(3,C)$ gauge transformation $\delta_{\vec{\eta}}$, the connection $A^a_{\mu}$ transforms as \cite{GRAV}
\begin{eqnarray}
\label{COMPLETENESS}
\delta_{\vec{\eta}}A^a_{\mu}=-D_{\mu}\eta^a=-\partial_{\mu}\eta^a-f^{abc}A^b_{\mu}\eta^c,
\end{eqnarray}
\noindent
which is the covariant derivative of the gauge parameter with $SO(3,C)$ structure constants $f^{abc}$.  Under infinitesimal spacetime diffeomorphisms $\delta_{\xi}$, the connection $A^a_{\mu}$ transforms according to the Lie derivative \cite{GRAV}
\begin{eqnarray}
\label{COMPLETENESS5}
\delta_{\xi}A^a_{\mu}=\xi^{\nu}\partial_{\nu}A^a_{\mu}+(\partial_{\mu}\xi^{\nu})A^a_{\nu}.
\end{eqnarray}
\noindent
We will show that transformations (\ref{COMPLETENESS}) and (\ref{COMPLETENESS5}) form a symmetry group of $A^a_{\mu}$.  More precisely, we will show without using any equations of motion or Poisson brackets that the Lie bracket of these transformations acting on $A^a_{\mu}$ yields a linear combination of the same transformations acting on $A^a_{\mu}$.  Let us quote here the final results of the algebra which will be computed in this section.  The final results of the $SO(3,C)*Diff$ algebra are
\begin{eqnarray}
\label{FINALRESULTS}
\bigl[\delta_{\vec{\theta}},\delta_{\vec{\eta}}\bigr]A^a_{\mu}=-\delta_{\vec{\theta}\times\vec{\eta}}A^a_{\mu};~~\bigl[\delta_{\xi},\delta_{\vec{\eta}}\bigr]A^a_{\mu}=-\delta_{(\delta_{\xi},\vec{\eta})}A^a_{\mu};~~
\bigl[\delta_{\xi},\delta_{\zeta}\bigr]A^a_{\mu}=-\delta_{[\xi,\zeta]}A^a_{\mu},
\end{eqnarray}
\noindent
namely that $SO(3,C)*Diff$ is a symmetry group of a $SO(3,C)$ gauge connection $A^a_{\mu}$.  The remaining subsections will prove that this is indeed the case by explicit calculation.

\subsection{The subalgebra of $SO(3,C)$ gauge transformations}

First we will show that the algebra of $SO(3,C)$ gauge transformations closes on this field.
Acting on (\ref{COMPLETENESS}) with a second gauge transformation $\delta_{\vec{\theta}}$ and using (\ref{COMPLETENESS}), we have
\begin{eqnarray}
\label{COMPLETENESS1}
\delta_{\vec{\theta}}\delta_{\vec{\eta}}A^a_{\mu}=-f^{abc}(\delta_{\vec{\theta}}A^b_{\mu})\eta^c=-f^{abc}(D_{\mu}\theta^b)\eta^c\nonumber\\
=f^{abc}\eta^c\partial_{\mu}\theta^b+\eta^c\bigl(\delta^{cf}\delta^{ag}-\delta^{cg}\delta^{af}\bigr)A^f_{\mu}\theta^g.
\end{eqnarray}
\noindent
We must then compute the result with the transformations reversed
\begin{eqnarray}
\label{COMM}
\delta_{\vec{\theta}}\delta_{\vec{\eta}}A^a_{\mu}=f^{abc}\eta^c\partial_{\mu}\theta^b+A^f_{\mu}\eta^f\theta^a-A^a_{\mu}(\vec{\eta}\cdot\vec{\theta});\nonumber\\
\delta_{\vec{\eta}}\delta_{\vec{\theta}}A^a_{\mu}=f^{abc}\theta^c\partial_{\mu}\eta^b+A^f_{\mu}\theta^f\eta^a-A^a_{\mu}(\vec{\theta}\cdot\vec{\eta}).
\end{eqnarray}
\noindent
Subtraction of the bottom line of (\ref{COMM}) from the top line yields the commutator of the two transformations
\begin{eqnarray}
\label{COMPLETENESS2}
\bigl[\delta_{\vec{\theta}},\delta_{\vec{\eta}}\bigr]=\partial_{\mu}(f^{abc}\theta^b\eta^c)+f^{afc}A^f_{\mu}(f^{cge}\theta^g\eta^e)=D_{\mu}(\vec{\theta}\times\vec{\eta})^a,
\end{eqnarray}
\noindent
which one recognizes as a $SO(3,C)$ gauge transformation with composite parameter $f^{abc}\theta^b\eta^c$.  Therefore the result is 
\begin{eqnarray}
\label{COMPLETENESS4}
\bigl[\delta_{\vec{\theta}},\delta_{\vec{\eta}}\bigr]A^a_{\mu}=-\delta_{\vec{\theta}\times\vec{\eta}}A^a_{\mu},
\end{eqnarray}
\noindent
namely that the commutator of two $SO(3,C)$ gauge transformations acting on $A^a_{\mu}$ is a $SO(3,C)$ gauge transformation acting on $A^a_{\mu}$.  Therefore, the connection $A^a_{\mu}$ provides a realization of $SO(3,C)$ gauge transformations, seen as a symmetry group.\par
\indent

\subsection{Mixed components of the algebra}

Next we move to the commutator of a gauge transformation with a spacetime diffeomorphism.  The variation by diffeomorphism of the variation by $SO(3,C)$ gauge transformation of $A^a_{\mu}$ is given by
\begin{eqnarray}
\label{COMPLETENESS6}
\delta_{\vec{\xi}}(\delta_{\vec{\eta}}A^a_{\mu})=-\delta_{\xi}\bigl(\partial_{\mu}\eta^a+f^{abc}A^b_{\mu}\eta^c\bigr)=-f^{abc}\eta^c(\delta_{\xi}A^b_{\mu})\nonumber\\
=-f^{abc}\eta^c\bigl(\xi^{\nu}\partial_{\nu}A^b_{\mu}+(\partial_{\mu}\xi^{\nu})A^b_{\nu}\bigr),
\end{eqnarray}
\noindent
where we have used (\ref{COMPLETENESS}) and (\ref{COMPLETENESS5}).  Reversing the order of the transformations yields, after making use of (\ref{COMPLETENESS}) and (\ref{COMPLETENESS5}),
\begin{eqnarray}
\label{COMPLETENESS7}
\delta_{\vec{\eta}}(\delta_{\xi}A^a_{\mu})=\xi^{\nu}\partial_{\nu}(\delta_{\vec{\eta}}A^a_{\mu})+(\partial_{\mu}\xi^{\nu})(\delta_{\vec{\eta}}A^a_{\nu})\nonumber\\
=-\xi^{\nu}\partial_{\nu}\bigl(\partial_{\mu}\eta^a+f^{abc}A^b_{\mu}\eta^c\bigr)-(\partial_{\mu}\xi^{\nu})\bigl(\partial_{\nu}\eta^a+f^{abc}A^b_{\nu}\eta^c\bigr).
\end{eqnarray}
\noindent
We subtract (\ref{COMPLETENESS7}) from (\ref{COMPLETENESS6}) to get the commutator of the transformations, which is
\begin{eqnarray}
\label{COMPLETENESS8}
\bigl[\delta_{\xi},\delta_{\vec{\eta}}\bigr]A^a_{\mu}=\xi^{\nu}\partial_{\nu}\partial_{\mu}\eta^a+f^{abc}\xi^{\nu}(\partial_{\nu}A^b_{\mu})\eta^c-f^{abc}\eta^c\xi^{\nu}(\partial_{\nu}A^b_{\mu})\nonumber\\
+f^{abc}\xi^{\nu}A^b_{\mu}(\partial_{\nu}\eta^c)-f^{abc}\eta^c(\partial_{\mu}\xi^{\nu})A^b_{\nu}+(\partial_{\mu}\xi^{\nu})(\partial_{\nu}\eta^a)+f^{abc}(\partial_{\mu}\xi^{\nu})A^b_{\nu}\eta^c.
\end{eqnarray}
\noindent
The second, third, fifth and seventh terms on the right hand side of (\ref{COMPLETENESS8}) cancel out.  After applying the Liebniz rule to the first term of (\ref{COMPLETENESS8}), we are left with
\begin{eqnarray}
\label{COMPLETENESS9}
\partial_{\mu}(\xi^{\nu}\partial_{\nu}\eta^a)-(\partial_{\mu}\xi^{\nu})(\partial_{\nu}\eta^a)\nonumber\\
+f^{abc}A^b_{\mu}(\xi^{\nu}\partial_{\nu}\eta^c)+(\partial_{\mu}\xi^{\nu})(\partial_{\nu}\eta^a)=D_{\mu}(\xi^{\nu}\partial_{\nu}\eta^a).
\end{eqnarray}
\noindent
So the final result is 
\begin{eqnarray}
\label{COMPLETENESS10}
\bigl[\delta_{\xi},\delta_{\vec{\eta}}\bigr]A^a_{\mu}=-\delta_{[\delta_{\xi},\vec{\eta}]}A^a_{\mu},
\end{eqnarray}
\noindent
namely that the commutator of a spacetime diffeomorphism with a $SO(3,C)$ gauge transformation acting on $A^a_{\mu}$ is a $SO(3,C)$ gauge transformation acting on $A^a_{\mu}$.\par
\indent
  
\subsection{The subalgebra of spacetime diffeomorphisms}

Lastly, we must check that spacetime diffeomorphisms form a subalgebra of $SO(3,C)*Diff$ on the connection $A^a_{\mu}$.  A diffeomorphism parametrized by $\zeta$ followed by a diffeomorphism parametrized by $\xi$ is given by
\begin{eqnarray}
\label{COMPLETENESS11}
\delta_{\xi}\delta_{\zeta}A^a_{\mu}=\zeta^{\nu}\partial_{\nu}(\delta_{\xi}A^a_{\mu})+(\partial_{\mu}\zeta^a)(\delta_{\xi}A^a_{\nu})\nonumber\\
=\zeta^{\nu}\partial_{\nu}\bigl(\xi^{\sigma}\partial_{\sigma}A^a_{\mu}+(\partial_{\mu}\xi^{\sigma})A^a_{\sigma}\bigr)
+(\partial_{\mu}\zeta^{\nu})\bigl(\xi^{\sigma}\partial_{\sigma}A^a_{\nu}+(\partial_{\nu}\xi^{\sigma})A^a_{\sigma}\bigr)\nonumber\\
=\zeta^{\nu}(\partial_{\nu}\xi^{\sigma})\partial_{\sigma}A^a_{\mu}+\zeta^{\nu}\xi^{\sigma}\partial_{\nu}\partial_{\sigma}A^a_{\mu}+\zeta^{\nu}(\partial_{\nu}\partial_{\mu}\xi^{\sigma})A^a_{\sigma}\nonumber\\
+\zeta^{\nu}(\partial_{\mu}\xi^{\sigma})\partial_{\nu}A^a_{\sigma}+(\partial_{\mu}\zeta^{\nu})\xi^{\sigma}\partial_{\sigma}A^a_{\nu}+(\partial_{\mu}\zeta^{\nu})(\partial_{\nu}\xi^{\sigma})A^a_{\sigma}.
\end{eqnarray}
\noindent
The result with the vector fields interchanged is given by
\begin{eqnarray}
\label{COMPLETENESS12}
\delta_{\zeta}\delta_{\xi}A^a_{\mu}
=\xi^{\nu}(\partial_{\nu}\zeta^{\sigma})\partial_{\sigma}A^a_{\mu}+\xi^{\nu}\zeta^{\sigma}\partial_{\nu}\partial_{\sigma}A^a_{\mu}+\xi^{\nu}(\partial_{\nu}\partial_{\mu}\zeta^{\sigma})A^a_{\sigma}\nonumber\\
+\xi^{\nu}(\partial_{\mu}\zeta^{\sigma})\partial_{\nu}A^a_{\sigma}+(\partial_{\mu}\xi^{\nu})\zeta^{\sigma}\partial_{\sigma}A^a_{\nu}+(\partial_{\mu}\xi^{\nu})(\partial_{\nu}\zeta^{\sigma})A^a_{\sigma}.
\end{eqnarray}
\noindent
Subtracting (\ref{COMPLETENESS12}) from (\ref{COMPLETENESS11}), on the right hand side the second terms directly cancel and the fourth and fifth terms cross-cancel.  Application of the Liebniz rule to the third term leads to some more cancellations
\begin{eqnarray}
\label{COMPLETENESS13}
\bigl[\delta_{\xi},\delta_{\zeta}\bigr]A^a_{\mu}=\bigl(\zeta^{\nu}\partial_{\nu}\xi^{\sigma}-\xi^{\nu}\partial_{\nu}\zeta^{\sigma}\bigr)\partial_{\sigma}A^a_{\mu}\nonumber\\
+\partial_{\mu}\bigl(\zeta^{\nu}\partial_{\nu}\xi^{\sigma}-\xi^{\nu}\partial_{\nu}\zeta^{\sigma}\bigr)A^a_{\sigma}\nonumber\\
-(\partial_{\mu}\zeta^{\nu})(\partial_{\nu}\zeta^{\sigma})A^a_{\sigma}+(\partial_{\mu}\zeta^{\nu})(\partial_{\nu}\xi^{\sigma})A^a_{\sigma}
+(\partial_{\mu}\xi^{\nu})(\partial_{\nu}\zeta^{\sigma})A^a_{\sigma}-(\partial_{\mu}\xi^{\nu})(\partial_{\nu}\zeta^{\sigma})A^a_{\sigma},
\end{eqnarray}
\noindent
which annihilate all four terms on the bottom line of (\ref{COMPLETENESS13}).  The result is is an off-shell closure of the algebra in congruity with (\ref{LIE3})
\begin{eqnarray}
\label{COMPARISON}
\bigl[\delta_{\xi},\delta_{\zeta}\bigr]A^a_{\mu}=-\delta_{[\xi,\zeta]}A^a_{\mu},
\end{eqnarray}
\noindent
which shows that spacetime diffeomorphisms are realizable as a symmetry group of the connection $A^a_{\mu}$.\par
\indent

\section{The Ashtekar action}

Having shown that the connection $A^a_{\mu}$ forms an off-shell representation of $SO(3,C)*Diff$ symmetry, let us see to what extent this can be implemented at the canonical level.  In the canonical formulation one performs a 3+1 splitting of the initially convariant dynamical variables into spatial and temporal parts.  For a prototype action based on the connection $A^a_{\mu}$ let us use the Ashtekar formulation of GR \cite{ASH1}, which is given by
\begin{eqnarray}
\label{ACTIONASH}
I_{Ash}=\int{dt}\int_{\Sigma}d^3x\Bigl[\widetilde{\sigma}^i_a\dot{A}^a_i+A^a_0G_a-N^iH_i-{i \over 2}\underline{N}H\Bigr].
\end{eqnarray}
\noindent
This is a totally constrained system with $(A^a_0,N^i,\underline{N})$, respectively the $SO(3,C)$ rotation angle $A^a_0$, the shift vector $N^i$ and the densitized lapse 
function $\underline{N}=N(\hbox{det}\widetilde{\sigma})^{-1/2}$ as auxilliary fields.  The phase space variables are the self-dual Ashtekar connection $A^a_i$ and the densitized 
triad $\widetilde{\sigma}^i_a$.  These variables define fundamental Poisson brackets
\begin{eqnarray}
\label{THESEDEFINE}
\{A^a_i(x,t),\widetilde{\sigma}^j_b(y,t)\}=\delta^a_n\delta^j_i\delta^{(3)}(x,y);~~\{A^a_i,A^b_j\}=\{\widetilde{\sigma}^i_a,\widetilde{\sigma}^j_b\}=0,
\end{eqnarray}
\noindent
which induce the following Poisson brackets between any two smooth phase space functions $f$ and $g$ given by
\begin{eqnarray}
\label{THESEDEFINEONE}
\{f,g\}=\int_{\Sigma}d^3y\Bigl[{{\delta{f}} \over {\delta{A}^a_i(y)}}{{\delta{g}} \over {\delta{\widetilde{\sigma}^i_a(y)}}}
-{{\delta{g}} \over {\delta{A}^a_i(y)}}{{\delta{f}} \over {\delta{\widetilde{\sigma}^i_a(y)}}}\Bigr].
\end{eqnarray}
\noindent
The constraints in (\ref{ACTIONASH}) smearing the auxilliary fields are $(G_a,H_i,H)$ the Gauss' law, vector and Hamiltonian constraints
\begin{eqnarray}
\label{ACTIONASH1}
G_a=D_i\widetilde{\sigma}^i_a;~~H_i=\epsilon_{ijk}\widetilde{\sigma}^j_aB^k_a;~~H=\epsilon_{ijk}\epsilon^{abc}\widetilde{\sigma}^i_a\widetilde{\sigma}^j_b\Bigl({\Lambda \over 3}\widetilde{\sigma}^k_c+B^k_c\Bigr).
\end{eqnarray}
\noindent
Choosing $g$ in (\ref{THESEDEFINEONE}) to be one of the constraints (\ref{ACTIONASH1}) with $f$ as one of the dynamical variables, one can determine the variation of the latter under the transformations generated by the constraints.  
Under transformations generated by the Gauss' law constraint, the `small' gauge transformations, the Ashtekar variables transform as
\begin{eqnarray}
\label{GAUGEE}
\delta_{\vec{\eta}}A^a_i=-D_i\eta^a;~~\delta_{\vec{\eta}}\widetilde{\sigma}^i_a=-f_{abc}\widetilde{\sigma}^i_b\eta^c.
\end{eqnarray}
\noindent
Comparison of (\ref{GAUGEE}) with (\ref{COMPLETENESS}) shows at least that the canonical structure of (\ref{ACTIONASH}) preserves the correct transformation properties of the spatial part of $A^a_{\mu}$.  The magnetic 
field $B^i_a$ should transform as a covariant $SO(3,C)$ vector under gauge transformations
\begin{eqnarray}
\label{CINEMA}
\delta_{\vec{\eta}}B^i_e=\epsilon^{ijk}D_j(\delta{A}^e_k)=-\epsilon^{ijk}D_jD_k\eta^e=-f_{efg}B^i_f\eta^g,
\end{eqnarray}
\noindent
which is indeed the case as expected.  Under transformations generated by the vector constraint, the Ashtekar variables transform as
\begin{eqnarray}  
\label{THEASH}
\delta_{\vec{N}}A^a_i=\epsilon_{ijk}B^j_aN^k;~~\delta_{\vec{N}}\widetilde{\sigma}^i_a=\widetilde{\sigma}^j_a\partial_jN^i-\partial_j(N^j\widetilde{\sigma}^i_a)-f_{abc}(N^jA^b_j)\widetilde{\sigma}^i_c.
\end{eqnarray}
\noindent
Comparison of (\ref{THEASH}) with (\ref{COMPARISON}) shows that the Ashtekar vector constraint does not generate pure spatial diffeomorphisms, but also includes a gauge transformation with field-dependent parameter $N^iA^a_i$.  So far, the spatial components of the Ashtekar variables exhibit the expected transformation properties with respect to $diff$ implied by the Poisson brackets (\ref{THESEDEFINEONE}).  This leaves remaining the `temporal' diffeomorphisms, which can be seen as the Hamilton's equations of motion.\par
\indent
The evolution equations of the Ashtekar variables generated purely by the Hamiltonian constraint are given by
\begin{eqnarray}
\label{EVOLUTION}
\delta_NA^a_i=-i\underline{N}\bigl(\Lambda(\hbox{det}\widetilde{\sigma})(\widetilde{\sigma}^{-1})^a_i+\epsilon_{ijk}\epsilon^{abc}\widetilde{\sigma}^j_bB^k_c\bigr);\nonumber\\
\delta_N\widetilde{\sigma}^i_a=-iD^{ni}_{ca}\bigl(N\sqrt{\hbox{det}\widetilde{\sigma}}(\widetilde{\sigma}^{-1})^c_n\bigr).
\end{eqnarray}
\noindent
The connection has acquired momentum space dependence under Hamiltonian evolution, which is no longer consistent with (\ref{COMPLETENESS5}).  We say that the 4-dimensional diffeomorphism symmetry is no longer manifestly preserved by the Poisson brackets (\ref{THESEDEFINEONE}).  The constraints algebra for (\ref{ACTIONASH1}) reads \cite{ASH2}
\begin{eqnarray}
\label{ALGEBRA12}
\{\vec{H}[\vec{N}],\vec{H}[\vec{M}]\}=H_k\bigl[N^{i}\partial^{k}M_i-M^{i}\partial^{k}N_i\bigr];\nonumber\\
\{\vec{H}[N],G_a[\theta^a]\}=G_a[N^{i}\partial_{i}\theta^a];\nonumber\\
\{G_a[\theta^a],G_b[\lambda^b]\}=G_{a}\bigl[f^a_{bc}\theta^{b}\lambda^c\bigr];\nonumber\\
\{H[\underline{N}],\vec{H}[\vec{N}]\}=H[N^{i}\partial_{i}\underline{N}\bigr]\nonumber\\
\{H[\underline{N}],G_a(\theta^a)\}=0;\nonumber\\
\bigl[H(\underline{N}),H[\underline{M}]\bigr]
=H_{i}[\bigl(\underline{N}\partial_{j}\underline{M}
-\underline{M}\partial_{j}\underline{N}\bigr)H^{ij}]
\end{eqnarray}
\noindent
which closes.  The algebra (\ref{ALGEBRA12}) is first class, and therefore the theory based on (\ref{ACTIONASH}) is in this sense Dirac consistent.\par
\indent
Still, comparison of (\ref{ALGEBRA12}) with (\ref{LIE}) reveals the same differences pointed out in the introduction, regarding the temporal parts of spacetime diffeomorphisms.  In particular, the bracket between two Hamiltonian constraints is a diffeomorphism (and not a Hamiltonian) constraint.  The suggestion is that that for the theory (\ref{ACTIONASH}), spacetime covariance at the canonical level has either been broken or is not manifest.  This leaves open the question of whether a formulation of GR exists where two Hamiltonian constraints commute into a Hamiltonian constraint as suggested by (\ref{LIE5}).\footnote{We will provide a proposition of such an action in the discussion section, relegating its comparison against (\ref{LIE5}) for future research.}\par

\subsection{Comparison with the covariant form}

Comparison of (\ref{ALGEBRA12}) with (\ref{FINALRESULTS}) with  reveals the aforementioned differences with respect to the temporal components of $Diff$ as shown in the introduction.  This suggests that the manifest covariance has been broken, since (\ref{ACTIONASH}) at the canonical level does not exhibit the same symmetry as the basic fields $A^a_{\mu}$.  We will nevertheless still complete the 
analysis of (\ref{FINALRESULTS}) with respect to gauge transformations for purposes of comparison with (\ref{ALGEBRA12}).  From (\ref{COMPLETENESS10}) the commutator of two gauge transformations is given by
\begin{eqnarray}
\label{LIE2}
\bigl[\delta_{\vec{\eta}},\delta_{\vec{\theta}}\bigr]=-\delta_{[\vec{\eta},\vec{\theta}]},
\end{eqnarray}
\noindent
which is consistent with (\ref{ALGEBRA12}).  For the result of a spacetime diffeomorphism with a gauge transformation we must revert to (\ref{COMPLETENESS10})
\begin{eqnarray}
\label{LIE1}
\bigl[\xi^{\mu}\partial_{\mu},\delta_{\vec{\theta}}\bigr]=-\delta_{L_{\xi}\theta}.
\end{eqnarray}
\noindent
\noindent
To compare (\ref{LIE1}) with (\ref{ALGEBRA12}) we must decompose the spacetime diffeomorphisms into a purely spatial part and a purely temporal part.  The commutator with the spatial part is given by
\begin{eqnarray}
\label{LIE6}
\bigl[N^i\partial_i,\delta_{\vec{\theta}}\bigr]=-\delta_{L_N\vec{\theta}},
\end{eqnarray}
\noindent
which is consistent with (\ref{ALGEBRA12}).  This shows that the Lie algebra of $SO(3,C)*diff$ is correctly realized by (\ref{ACTIONASH}) at the canonical level.  For the commutator of 
a gauge transformation and a temporal diffeomorphism (\ref{LIE1}) yields
\begin{eqnarray}
\label{LIE7}
\bigl[N\partial_0,\delta_{\vec{\theta}}\bigr]=-\delta_{N\dot{\theta}},
\end{eqnarray}
\noindent
which is consistent with (\ref{ALGEBRA12}) only for gauge transformations which are independent of time.  This is quite analogous to (\ref{LIE4}) in the introduction.\par
\indent

\section{Transformation properties of the CDJ matrix}
  
We have seen, while the 4-D connection $A^a_{\mu}$ forms a representation of $SO(3)*Diff$ symmetry, that only the spatial part of this symmetry is preserved at the canonical level in the Ashtekar formalism (\ref{ACTIONASH}) which uses $A^a_{\mu}$ as one of the dynamical variables.  We will introduce a new quantity $\Psi_{ae}\in{SO}(3,C)\otimes{SO}(3,C)$, which in conjunction with $A^a_{\mu}$ will form the dynamical variables for the instanton representation of GR which we will propose in the discussion section.  The matrix $\Psi_{ae}$ is known as the CDJ matrix, named after Capovilla, Dell and Jacobson \cite{NOMETRIC}.\par
\indent
Consider the following change of variables
\begin{eqnarray}
\label{CINEMA1}
\widetilde{\sigma}^i_a=\Psi_{ae}B^i_e;~(\hbox{det}\Psi)\neq{0},~(\hbox{det}B)\neq{0}.
\end{eqnarray}
We will verify that $\Psi_{ae}$ forms an off-shell representation of the $SO(3,C)*diff$ algebra, relegating the temporal part as a direction of future research.  First, let us determine the transformation properties of $\Psi_{ae}$ predicted by this algebra.  The connection $A^a_i$ and densitized 
triad $\widetilde{\sigma}^i_a$ under spatial diffeomorphisms $diff$ generated by the vector $N^i$ should transform respectively as covariant and contravariant 3-vectors
\begin{eqnarray}
\label{CINEMA5}
\delta_{\vec{N}}A^a_i=N^j\partial_jA^a_i+(\partial_iN^j)A^a_j;~~\delta_{\vec{N}}\widetilde{\sigma}^j_e=N^k\partial_k\widetilde{\sigma}^j_e-\widetilde{\sigma}^i_e(\partial_iN^j).
\end{eqnarray}
\noindent
The first equation of (\ref{CINEMA5}) can also be written as 
\begin{eqnarray}
\label{CINEMA6}
\delta_{\vec{N}}A^a_k=D_k(N^mA^a_m)+\epsilon_{klm}B^l_aN^m.
\end{eqnarray}
\noindent
Comparison of (\ref{CINEMA6}) and the second equation of (\ref{CINEMA5}) with (\ref{THEASH}) shows that the Ashtekar vector constraint generates spatial diffeomorphisms combined with a gauge transformation with field-dependent parameter $N^iA^a_i$.  Under $diff$, the magentic field $B^i_a$ transforms as 
\begin{eqnarray}
\label{CINEMA7}
\delta_{\vec{N}}B^i_a=\epsilon^{ijk}D_j(\delta{A}^a_k)
=\epsilon^{ijk}D_jD_k(N^mA^a_m)+\epsilon^{ijk}\epsilon_{lmk}D_j(N^mB^l_a)\nonumber\\
=f^{abc}B^i_b(N^mA^c_m)+\bigl(\delta^i_l\delta^j_m-\delta^i_m\delta^j_l\bigr)D_j(N^mB^l_a),
\end{eqnarray}
\noindent
were we have used (\ref{CINEMA6}), the relation $\delta{B}^i_a=\epsilon^{ijk}D_j(\delta{A}^a_k)$, as well as the definition of curvature as the commutator of two covariant derivatives.  Continuing from (\ref{CINEMA7}), we have
\begin{eqnarray}
\label{CINEMA8}
\delta_{\vec{N}}B^i_a
=f^{abc}B^i_b(N^mA^c_m)+(\partial_mN^m)B^i_a\nonumber\\
+N^m\partial_mB^i_a+N^mf_{abc}A^b_mB^i_c-(\partial_lN^i)B^l_a-N^iD_lB^l_a.
\end{eqnarray}
\noindent
Using the Bianchi identity $D_lB^l_a=0$ and the fact that the first and fourth terms on the right hand side of (\ref{CINEMA8}) cancel, we are left with
\begin{eqnarray}
\label{CINEMA9}
\delta_{\vec{N}}B^i_a=\partial_m(N^mB^i_a)-(\partial_lN^i)B^l_a.
\end{eqnarray}
\noindent 
Comparison with the second equation of (\ref{CINEMA5}) shows up to a term proportional to $\partial_mN^m$ that equation (\ref{CINEMA9}) is consistent with the transformation properties expected of a contravariant 3-vector.\par
\indent
Having written down explicitly the transformations of the Ashtekar variables under $SO(3,C)*diff$, we will now use this to deduce the transformation properties 
of $\Psi_{ae}$ using (\ref{CINEMA1}).  Starting with the $SO(3,C)$ gauge transformations, the variation of (\ref{CINEMA1}) is given by
\begin{eqnarray}
\label{CINEMA2}
\delta_{\vec{\eta}}\widetilde{\sigma}^i_a=\Psi_{ae}(\delta_{\vec{\eta}}{B}^i_e)+(\delta_{\vec{\eta}}\Psi_{ae})B^i_e,
\end{eqnarray}
\noindent
where we have used the Liebniz rule.  Using (\ref{CINEMA}), this yields the condition
\begin{eqnarray}
\label{CINEMA3}
-f_{abc}\widetilde{\sigma}^i_b\eta^c=\Psi_{ae}(-f_{ebc}B^i_b\eta^c)+(\delta_{\vec{\eta}}\Psi_{ae})B^i_e,
\end{eqnarray}
\noindent
where we have used the Liebniz rule.  Using (\ref{CINEMA}) and (\ref{CINEMA9}) we have after relabelling indices $e\leftrightarrow{g}$ and $b\leftrightarrow{e}$ on the second term
\begin{eqnarray}
\label{CINEMA31}
-f_{abc}\Psi_{be}\eta^cB^i_e=-\Psi_{ag}f_{gec}B^i_e\eta^c+(\delta_{\vec{\eta}}\Psi_{ae})B^i_e.
\end{eqnarray}
\noindent
Assuming the nondegeneracy condition for $B^i_e$, then (\ref{CINEMA31}) gives us the transformation properties of $\Psi_{ae}$
\begin{eqnarray}
\label{CINEMA4}
\delta_{\vec{\eta}}\Psi_{ae}=-\bigl(f_{abc}\Psi_{be}+\Psi_{ab}f_{ebc}\bigr)\eta^c.
\end{eqnarray}
\noindent
The result is that $\Psi_{ae}$ transforms as a second-rank $SO(3,C)$ tensor under gauge transformations, which seems to makes sense.  Next, we would like to determine the transformation properties of $\Psi_{ae}$ under spatial diffeomorphisms.  Note that we have  
\begin{eqnarray}
\label{CINEMA10}
\delta_{\vec{N}}\widetilde{\sigma}^i_a=\Psi_{ae}(\delta_{\vec{N}}{B}^i_e)+(\delta_{\vec{N}}\Psi_{ae})B^i_e,
\end{eqnarray}
\noindent
the variation of (\ref{CINEMA1}) under spatial diffeomorphisms.  Using (\ref{CINEMA5}), we have
\begin{eqnarray}
\label{CINEMA11}
N^m\partial_m\widetilde{\sigma}^i_a-(\partial_mN^i)\widetilde{\sigma}^m_a
=(\delta_{\vec{N}}\Psi_{ae})B^i_e+\Psi_{ae}\partial_m(N^mB^i_e)-(\partial_mN^i)\Psi_{ae}B^m_e.
\end{eqnarray}
\noindent
The second term on the left hand side of (\ref{CINEMA11}) cancels the last term on the right hand side, yielding
\begin{eqnarray}
\label{CINEMA12}
N^m\partial_m\widetilde{\sigma}^i_a=(\delta_{\vec{N}}\Psi_{ae})B^i_e+\Psi_{ae}\partial_m(N^mB^i_e).
\end{eqnarray}
\noindent
Substituting (\ref{CINEMA1}) and expanding the derivatives, we have
\begin{eqnarray}
\label{CINEMA13}
N^m\partial_m(\Psi_{ae}B^i_e)=(\delta_{\vec{N}}\Psi_{ae})B^i_e+\Psi_{ae}\partial_m(N^mB^i_e)\nonumber\\
\longrightarrow
N^m(\partial_m\Psi_{ae})B^i_e+N^m\Psi_{ae}\partial_mB^i_e\nonumber\\
=(\delta_{\vec{N}}\Psi_{ae})B^i_e+\Psi_{ae}(\partial_mN^m)B^i_e+\Psi_{ae}N^m(\partial_mB^i_e).
\end{eqnarray}
\noindent
After cancellation of various terms in (\ref{CINEMA13}), we obtain
\begin{eqnarray}
\label{CINEMA14}
\delta_{\vec{N}}\Psi_{ae}=N^m\partial_m\Psi_{ae}-(\partial_mN^m)\Psi_{ae}.
\end{eqnarray}
\noindent
Equations (\ref{CINEMA4}) and (\ref{CINEMA14}) provide the desired transformation properties of $\Psi_{ae}$ under $SO(3,C)*diff$.  We will now prove that $\Psi_{ae}$ forms an off-shell representation of the $SO(3,C)*diff$ algebra, which we have relegated to the appendix.  The result is shown in the discussion section.\par
\indent

\section{Discussion}

This paper has proposed variables $(A^a_{\mu},\Psi_{ae})$, with a view toward restoring the full covariance of general relativity under spacetime diffeomorphisms.  The connection $A^a_{\mu}$ forms an off-shell representation 
of $SO(3,C)*Diff$ symmetry, and in the case of $\Psi_{ae}$ we have proven only $SO(3,C)*diff$ symmetry, which does not include the temporal components of $Diff$.  The algebra is given by
\begin{eqnarray}
\label{WAVEITER}
\bigl[\delta_{\vec{\theta}},\delta_{\vec{\eta}}\bigr]A^a_{\mu}=-\delta_{\vec{\theta}\times\vec{\eta}}A^a_{\mu};~~
\bigl[\delta_{\vec{\theta}},\delta_{\vec{\eta}}\bigr]\Psi_{ae}=-\delta_{\vec{\theta}\times\vec{\eta}}\Psi_{ae}\nonumber\\
\bigl[\delta_{\xi},\delta_{\vec{\eta}}\bigr]A^a_{\mu}=-\delta_{\delta_{L_{\xi}}\vec{\eta}}A^a_{\mu};~~
\bigl[\delta_{\vec{N}},\delta_{\vec{\eta}}\bigr]\Psi_{ae}=-\delta_{L_{\vec{N}}\vec{\eta}}\Psi_{ae}\nonumber\\
\bigl[\delta_{\xi},\delta_{\zeta}\bigr]A^a_{\mu}=-\delta_{[\xi,\zeta]}A^a_{\mu};~~
\bigl[\delta_{\vec{M}},\delta_{\vec{N}}\bigr]\Psi_{ae}=-\delta_{[\vec{N},\vec{M}]}\Psi_{ae}.
\end{eqnarray}
\noindent
We have shown that the Ashtekar formulation of general relativity defined on the phase space $\Omega_{Ash}=(\widetilde{\sigma}^i_a,A^a_i)$ implements only the $SO(3,C)*diff$ part of the algebra, and only for spatial diffeomorphisms and $SO(3,C)$ gauge transformations which are independent of time.  This suggests that the $SO(3,C)*Diff$ symmetry of the connection $A^a_{\mu}$ has been broken at the canonical level.  A question which we propose for future research is whether this symmetry can to some extent be restored at the canonical level using an alternate formulation of GR.  We will propose a new action 
\begin{eqnarray}
\label{ACTIONINSTTT}
I_{Inst}=\int{dt}\int_{\Sigma}d^3x\Bigl(\Psi_{ae}B^i_e\dot{A}^a_i+A^a_0B^i_eD_i\Psi_{ae}\nonumber\\
+\epsilon_{ijk}N^iB^j_aB^k_e\Psi_{ae}-iN(\hbox{det}B)^{1/2}\sqrt{\hbox{det}\Psi}\bigl(\Lambda+\hbox{tr}\Psi^{-1}\bigr)\Bigr),
\end{eqnarray}
\noindent
which can be obtained by substitution (\ref{CINEMA1}) into the action (\ref{ACTIONASH}).  Using the definition $F^a_{0i}=\dot{A}^a_i-D_iA^a_0$ for the temporal component of the curvature, and using the identity
\begin{eqnarray}
\label{TRANSFORMEDINTO}
{1 \over 2}\int{dt}\int_{\Sigma}\Psi_{(ae)}\epsilon^{ijk}F^e_{jk}F^a_{0i}={1 \over 8}\int_Md^4x\Psi_{ae}F^a_{\mu\nu}F^e_{\rho\sigma}\epsilon^{\mu\nu\rho\sigma},
\end{eqnarray}
\noindent
then upon separation of $\Psi_{ae}$ into symmetric and antisymmetric parts we can write the action (\ref{ACTIONINSTTT}) as
\begin{eqnarray}
\label{ACTIONINSTTT1}
I_{Inst}=\int{dt}\int_{\Sigma}d^3x\bigl(B^i_{[e}F^{a]}_{0i}+\epsilon_{ijk}N^iB^j_aB^k_e\bigr)\Psi_{ae}\nonumber\\
+\int_Md^4x\biggl[{1 \over 8}\Psi_{ae}F^a_{\mu\nu}F^e_{\rho\sigma}\epsilon^{\mu\nu\rho\sigma}-iN(\hbox{det}B)^{1/2}\sqrt{\hbox{det}\Psi}\bigl(\Lambda+\hbox{tr}\Psi^{-1}\bigr)\biggr].
\end{eqnarray}
\noindent
As a direction of future research we will check the algebra of constraints implied by (\ref{ACTIONINSTTT1}) to examine to what extent the $SO(3,C)*Diff$ symmetry can be restored.  Since (\ref{LIE5}) suggests that temporal diffeomorphisms should form their own subalgebra, an additional area of research will be to investigate whether this can be deduced from (\ref{ACTIONINSTTT}).  It is shown in \cite{EYOITA}, using a reduced version 
of (\ref{ACTIONINSTTT}), that the Hamiltonian constraint does form a subalgebra, albeit with structure functions and not structure constants.

\section{Appendix A: Closure of the $SO(3,C)*diff$ algebra on $\Psi_{ae}$}

\subsection{Closure under spatial diffeomorphisms}
Having determined the transformation properties of $\Psi_{ae}$, we will now check for off-shell closure of the constraints algebra on $\Psi_{ae}$, seen as a dynamical variable.  Let us first examine the effect of two consecutive diffeomorphisms
\begin{eqnarray}
\label{CINEMA15}
\delta_{\vec{M}}\delta_{\vec{N}}\Psi_{ae}
=N^m\partial_m(\delta_{\vec{M}}\Psi_{ae})-(\partial_mN^m)\delta_{\vec{M}}\Psi_{ae}\nonumber\\
=N^m\partial_m\Bigl(M^n\partial_n\Psi_{ae}-(\partial_nM^n)\Psi_{ae}\Bigr)
-(\partial_mN^m)\Bigl(M^n\partial_n\Psi_{ae}-(\partial_nM^n)\Psi_{ae}\Bigr)\nonumber\\
=N^m(\partial_mM^n)\partial_n\Psi_{ae}+N^mM^n\partial_m\partial_n\Psi_{ae}-(\partial_mN^m)M^n\partial_n\Psi_{ae}+(\partial_mN^m)(\partial_nM^n)\Psi_{ae}\nonumber\\
-N^m(\partial_m\partial_nM^n)\Psi_{ae}-N^m(\partial_nM^n)\partial_m\Psi_{ae},
\end{eqnarray}
\noindent
where we have used (\ref{CINEMA14}).  After cancellation of various terms and subtracting the result with $\vec{M}$ and $\vec{N}$ interchanged, we obtain
\begin{eqnarray}
\label{CINEMA16}
\bigl(\delta_{\vec{M}}\delta_{\vec{N}}-\delta_{\vec{N}}\delta_{\vec{M}}\bigr)\Psi_{ae}
=(L_{\vec{N}}\vec{M})^n\partial_n\Psi_{ae}+\bigl(M^n\partial_m\partial_nN^n-N^m\partial_m\partial_nM^n\bigr)\Psi_{ae}\nonumber\\
=(L_{\vec{N}}\vec{M})^n\partial_n\Psi_{ae}-\partial_n(L_{\vec{N}}\vec{M})^n\Psi_{ae}
\end{eqnarray}
\noindent
where $(L_{\vec{N}}\vec{M})^n=N^m\partial_mM^n-M^m\partial_mN^n$ is the $n^{th}$ component of the Lie derivative of $M^i$ along $N^i$.  The result is 
\begin{eqnarray}
\label{CINEMA17}
\bigl[\delta_{\vec{M}},\delta_{\vec{N}}\bigr]\Psi_{ae}=-\delta_{[\vec{N},\vec{M}]}\Psi_{ae},
\end{eqnarray}
\noindent
namely that the commutator of two spatial diffeomorphisms is also a spatial diffeomorphism.  The spatial diffeomorphisms form a closed algebra on $\Psi_{ae}$, which is consistent with (\ref{LIE3}) and with the purely spatial 
part of (\ref{ALGEBRA12}).\par
\indent

\subsection{Closure under mixed transformations}

Next we move on to the commutator of a spatial diffeomorphism with a gauge transformation.  Using (\ref{CINEMA4}) and (\ref{CINEMA14}), this is given by
\begin{eqnarray}
\label{CINEMA18}
\bigl[\delta_{\vec{N}},\delta_{\vec{\eta}}\bigr]\Psi_{ae}=\delta_{\vec{N}}(\delta_{\vec{\eta}}\Psi_{ae})-\delta_{\vec{\eta}}(\delta_{\vec{N}}\Psi_{ae})\nonumber\\
=-\bigl(f_{abc}\delta_{\vec{N}}\Psi_{be}+f_{ebc}\delta_{\vec{N}}\Psi_{ab}\bigr)\eta^c
-\bigl(N^m\partial_m(\delta_{\vec{\eta}}\Psi_{ae})-(\partial_mN^m)\delta_{\vec{\eta}}\Psi_{ae}\bigr)\nonumber\\
=-f_{abc}\bigl(N^m\partial_m\Psi_{be}-(\partial_mN^m)\Psi_{be}\bigr)\eta^c-f_{ebc}\bigl(N^m\partial_m\Psi_{ab}-(\partial_mN^m)\Psi_{ab}\bigr)\eta^c\nonumber\\
+\bigl(N^m\partial_m-(\partial_mN^m)\bigr)\bigl(f_{abc}\Psi_{be}+f_{ebc}\Psi_{ab}\bigr)\eta^c.
\end{eqnarray}
\noindent
Expanding the terms, we obtain
\begin{eqnarray}
\label{CINEMA19}
-f_{abc}N^m(\partial_m\Psi_{be})\eta^c+f_{abc}(\partial_mN^m)\Psi_{be}\eta^c-f_{ebc}N^m(\partial_m\Psi_{ab})\eta^c+f_{ebc}(\partial_mN^m)\Psi_{ab}\eta^c\nonumber\\
+f_{abc}N^m(\partial_m\Psi_{be})\eta^c+f_{abc}N^m\Psi_{be}(\partial_m\eta^c)+f_{ebc}N^m(\partial_m\Psi_{ab})\eta^c+f_{ebc}N^m\Psi_{ab}(\partial_m\eta^c)\nonumber\\
-f_{abc}(\partial_mN^m)\Psi_{be}\eta^c-f_{ebc}(\partial_mN^m)\Psi_{ab}\eta^c=(N^m\partial_m\eta^c)\bigl(f_{abc}\Psi_{be}+f_{ebc}\Psi_{ab}\bigr)
\end{eqnarray}
\noindent
which leads to cancellation of several terms.  The result is that
\begin{eqnarray}
\label{CINEMA20}
\bigl[\delta_{\vec{N}},\delta_{\vec{\eta}}\bigr]\Psi_{ae}=-\delta_{L_{\vec{N}}\vec{\eta}}\Psi_{ae},
\end{eqnarray}
\noindent
namely that the commutator of a gauge transformation and a spatial diffeomorphism is a gauge transformation, which is also consistent with (\ref{ALGEBRA12}).\par
\indent  

\subsection{Closure under $SO(3,C)$ gauge transformations}

Next we move on to the commutator of two gauge transformations.  Two successive gauge transformations yield
\begin{eqnarray}
\label{CINEMA21}
\delta_{\vec{\theta}}\delta_{\vec{\eta}}\Psi_{ae}=-\bigl(f_{abc}\delta_{\vec{\theta}}\Psi_{be}+f_{ebc}\delta_{\vec{\theta}}\Psi_{ab}\bigr)\eta^c.
\end{eqnarray}
\noindent
Expanding this out and using (\ref{CINEMA4}), we obtain
\begin{eqnarray}
\label{CINEMA22}
f_{abc}\bigl(f_{bfd}\Psi_{fe}+f_{efd}\Psi_{bf}\bigr)\theta^d\eta^c+f_{bec}\bigl(f_{afd}\Psi_{fb}+f_{bfd}\Psi_{af}\bigr)\theta^d\eta^c\nonumber\\
=\bigl(\delta_{ad}\delta_{cf}-\delta_{af}\delta_{cd}\bigr)\Psi_{fe}+f_{abc}f_{efd}\Psi_{bf}+f_{ebc}f_{afd}\Psi_{fb}
+\bigl(\delta_{ed}\delta_{cf}-\delta_{ef}\delta_{cd}\bigr)\Psi_{af}.
\end{eqnarray}
\noindent
Expanding (\ref{CINEMA22}) and subtracting the result with $\vec{\theta}$ and $\vec{\eta}$ interchanged, we obtain
\begin{eqnarray}
\label{CINEMA23}
\bigl[\delta_{\vec{\theta}},\delta_{\vec{\eta}}\bigr]\Psi_{ae}
=\Bigl(\delta_{ad}\Psi_{ce}-\delta_{cd}\Psi_{ae}+\delta_{ed}\Psi_{ac}-\delta_{cd}\Psi_{ae}+f_{abc}f_{efd}\Psi_{bf}+f_{ebc}f_{afd}\Psi_{fb}\Bigr)\theta^d\eta^c\nonumber\\
+\Bigl(-\delta_{ac}\Psi_{de}+\delta_{dc}\Psi_{ae}-\delta_{ec}\Psi_{ad}+\delta_{dc}\Psi_{ae}-f_{abd}f_{efc}\Psi_{bf}-f_{ebd}f_{afc}\Psi_{fb}\Bigr)\eta^c\theta^d
=A+B,
\end{eqnarray}
\noindent
where we have relabelled $c\leftrightarrow{d}$ in the second line.  We have also defined
\begin{eqnarray}
\label{CINEMA24}
A=\bigl(\delta_{ad}\Psi_{ce}-\delta_{ac}\Psi_{de}+\delta_{ed}\Psi_{ac}-\delta_{ec}\Psi_{ad}\bigr)\theta^d\eta^c;\nonumber\\
B=\Bigl(f_{abc}f_{efd}\Psi_{bf}+f_{ebc}f_{afd}\Psi_{fb}-f_{abd}f_{efc}\Psi_{bf}-f_{ebd}f_{afc}\Psi_{fb}\Bigr)\theta^d\eta^c.
\end{eqnarray}
\noindent
Note that $B=0$, which can easiest be seen by relabelling $f\leftrightarrow{b}$ on the second and fourth terms of (\ref{CINEMA24}).  Hence it remains to show that $A$ in (\ref{CINEMA4}) is a gauge transformation.  This can be written as 
\begin{eqnarray}
\label{CINEMA25}
A=\delta_{ad}\Psi_{ce}(\theta^d\eta^c-\theta^c\eta^d)+\delta_{ed}\Psi_{ac}(\theta^d\eta^c-\theta^c\eta^d)
=\bigl(\delta_{ad}\Psi_{ce}+\delta_{ed}\Psi_{ac}\bigr)(\theta^d\eta^c-\theta^c\eta^d).
\end{eqnarray}
\noindent
Defining $\theta^d\eta^c-\theta^c\eta^d=\epsilon^{dcg}q^g$, where $q^g=\epsilon^{ga^{\prime}e^{\prime}}\theta^{a^{\prime}}\eta^{e^{\prime}}$, then we have
\begin{eqnarray}
\label{CINEMA26}
A=(\delta_{ad}\Psi_{ce}+\delta_{ed}\Psi_{ac}\bigr)\epsilon_{dcg}q^g=\bigl(f_{acg}\Psi_{ce}+f_{ecg}\Psi_{ac}\bigr)q^g.
\end{eqnarray}
\noindent
The result is
\begin{eqnarray}
\label{CINEMA27}
\bigl[\delta_{\vec{\theta}},\delta_{\vec{\eta}}\bigr]\Psi_{ae}=-\delta_{\vec{\theta}\times\vec{\eta}}\Psi_{ae},
\end{eqnarray}
\noindent
namely that the commutator of two gauge transformations is a gauge transformation.  Comparison of the results of (\ref{CINEMA17}), (\ref{CINEMA20}) and (\ref{CINEMA27}) with (\ref{ALGEBRA12}) shows that the constraints algebra for kinematic transformations is preserved under the change of variables (\ref{CINEMA1}).  For the temporal part of the transformations of $\Psi_{ae}$, we will defer treatment to a separate paper.

\end{document}